\documentstyle[12pt]{article}
\begin{document}
\begin{titlepage}
\begin{flushright}
Report \#RU 93-3-B
\end{flushright}
\begin{center}
{\LARGE\bf Solvent-induced organization:\\
\vspace{2mm}
A physical model of folding myoglobin}\\
\end{center}

\vspace{7mm}
\begin{center}
{\bf David J.E. Callaway*\\
Department of Physics\\
The Rockefeller University\\
1230 York Avenue\\
New York, NY 10021-6399\\
USA}\\
callaway@physics.rockefeller.edu\\
\end{center}

\vspace{5mm}
\begin{abstract}
The essential features of the {\em in vitro} refolding
of myoglobin are expressed in a solvable
physical model.  Alpha helices are taken as the
fundamental collective coordinates of the system,
while the refolding is assumed to be mainly driven by
solvent-induced hydrophobic forces.  A quantitative
model of these forces is developed and compared
with experimental and theoretical results.
The model is then tested by being employed in a simulation scheme
designed to mimic solvent effects.  Realistic dynamic
trajectories of myoglobin are shown as it folds from an
extended conformation to a close approximation of the
native state.  Various suggestive features of the process
are discussed.  The tenets of the model are further
tested by folding the single-chain plant protein leghemoglobin.

\end{abstract}

\vspace{.50in}
\noindent
*Work supported in part by the U.S. Department of Energy under\\
Grant No. DOE-91ER40651 Task B.
\end{titlepage}

\newpage
\hsize=6in.
\hoffset=-.5in
\begin{center}
{\bf Solvent-induced organization:\\
A physical model of folding myoglobin}\\

\end{center}

\noindent
1. Introduction

\vspace{5mm}
a.  Synopsis

\vspace{5mm}
It is generally assumed that the native conformation of a
protein is given by the structure which yields the global
minimum of the free energy $^{[1]}$.  However, it is less
clear how a protein reaches this minimum. Simple counting
arguments $^{[2]}$ suggest that observed folding times,
typically 1-100 seconds, are far too short to allow a
global examination of all available states.  (This puzzle
is often referred to as the ``Levinthal paradox'').
Thus, there
are likely to be physical principles which determine the
way in which a protein folds.  Knowledge of these principles
should be of some utility in developing tools for predicting
protein folding.

\vspace{5mm}
A major stumbling block on the road to developing such tools
is the general absence of computationally tractable models
wherein a protein folds in a realistic fashion.  One candidate,
a model of the large-scale {\em in vitro} folding of myoglobin,
is presented here.  Two major assumptions about the folding
of myoglobin are made in this work and are discussed in
detail below.  First, it is assumed that the alpha helices
which comprise myoglobin are nascent at an early stage of
folding, and that the interactions between these larger
subunits determine the tertiary structure of this protein.
The second major assumption is that solvent effects (the
``hydrophobic'' interaction) are largely responsible
for this tertiary structure.

\vspace{5mm}
A quantitative model of hydrophobic forces
is developed below, using input from
experimental data.
It will be seen that the model also agrees
with microscopic models of effective hydrophobic interactions.
Molecules of the solvent water are included only implicitly
in the calculation.  This is in keeping with the reductionist
philosophy employed here, wherein the complexity of the
system is reduced by identifying and including only the
important degrees of freedom in the system.  Rather than
trying to fold a protein as precisely as possible, the object here
is to develop a tractable model, based upon sound physical principles,
which quickly allows one to determine global folding patterns
to ``sketchbook'' accuracy.
Uncertainties due to reparameterization of the potential
function are, accordingly, kept to a minimum.

\vspace{5mm}
More generic potential
functions could, of course, be included in
the present calculation to produce a more refined structure.
Such a generic approach, however, runs the risk of losing
contact with the physical principles which underly protein
folding.  This risk is lessened substantially if, rather than
postulating an intricate potential function and,
subsequently, determining
its parameters by fitting to a large data set, one instead
develops an explicit physical model for the relevant interaction.
By using realistic interaction potentials, the fundamental
laws which govern various aspects of protein folding can
be extracted systematically.  Additionally,
by refining the model through the elimination of nonessential interactions,
the computer time required to solve it is substantially reduced.
An extensive search of configuration space then becomes possible,
and the refined model then provides an excellent laboratory
for the testing and improvement of protein folding algorithms.

\vspace{5mm}
Following the specification of the model, it will be solved
using an approach based upon the Langevin equation.  This
formalism is particularly well-suited for approximating
solvent effects, for it describes a system interacting with
a medium which provides random thermal fluctuations.  When
the Langevin equations are integrated, the result is a
representative trajectory of the protein as it folds toward
its most probable configuration, the state of minimal free
energy.  A typical trajectory will be shown as a time series
of states leading to this minimum.  The resultant state of
minimum free energy forms a reasonable approximation to the
native conformation. The model is applied to two very
distinct proteins, sperm-whale myoglobin and leghemoglobin.
Despite the fact that these proteins differ greatly
in sequence and biological origin, they both
possess a similar topology (often called a ``globin fold'').
The existence of the globin fold and the topology of the
two proteins is accurately predicted by the model.

\newpage
\vspace{5mm}
b. Myoglobin -- an overview.

\vspace{5mm}
It has often been asserted that the tertiary structure of
myoglobin is a direct consequence of attractive interactions
between groups of hydrophobic residues on its constituent
alpha helices.  This assumption led to an early derivation
of the structure of myoglobin $^{[3]}$ by counting the
number of dehydrated regions for various configurations of
a simple geometrical model.  Subsequent work $^{[4]}$ was
based upon the assumption that helices pair at
specific hydrophobic interaction sites with certain helix
axis angles. This assumption reduces the allowed configuration space
considerably.  An initial set of 3 x 10$^8$ structures
is ultimately reduced to 20, one of which clearly resembles
myoglobin.  Important later work $^{[5]}$ posits a specific
form for the hydrophobic interaction between amino acid
residues, which are approximated as spheres of van der
Waals radius.  The force used is unphysically strong and
long-ranged and is the same between all hydrophobic
residues.  A minimization scheme is used to find
the configuration of minimum energy, which is similar to
native myoglobin.

\vspace{5mm}
Inspired by these successes, a physical model of
myoglobin is developed below and then solved
by a Langevin scheme designed to mimic solvent effects.
The model is intended to be part of a larger philosophy
which involves three stages of analysis.  In the first
stage, a tentative identification of regions of secondary
structure is made.  The second stage involves large-scale
movement of the tentatively assigned secondary structures,
which are taken as the fundamental degrees of freedom.
The third stage is a final ``polishing''
of the structure, using a detailed model of protein
dynamics (e.g., $^{[6]}$).  In principle, this final stage
employs a complete set of
realistic interactions and involves all degrees of freedom in the
model.  Due to the limitations of present computational
architecture, only small perturbations on the structures
generated by the large-scale global optimization of the second stage
can, therefore, be completed.  The first and third stages are
likely to be generally possible
in the near future; the second stage is contemplated
here.

\vspace{5mm}
Intrinsic to this approach is the idea that the folding
process can be described in terms of
``effective'' interactions between
fundamental units which are larger than atoms but
smaller than the entire protein $^{[65,66]}$.  One way this could
occur is via the collision and coalescence of unstable
microdomains, each of which folds and unfolds rapidly
(the ``diffusion-collision'' model $^{[7]}$).  The
existence of a stable folding intermediate is, therefore, not
necessary for the present approach to be valid.  Indeed,
extensive experimental work on refolding myoglobin
$^{[8-11]}$ generally finds no evidence of a such a folding
intermediate at normal (25-27$^o$C) temperatures.
(At -6$^o$C and 50$^o$C, however, folding intermediates
in apomyoglobin have been reported $^{[12,13]}$.)

\vspace{5mm}
\noindent
2.  Methods.

\vspace{5mm}
a.  Quantitative model of ``effective'' hydrophobic
forces.

\vspace{5mm}
The hydrophobic interaction is certainly a major component of
protein folding $^{[14]}$.  Moreover, the above-described
successes of models $^{[3-5]}$ for folding myoglobin suggest
that it is the dominant force between alpha helices in this
protein.  A quantitative physical
model for this force is now developed.

\vspace{5mm}
At normal temperatures, the hydrophobic interaction is
generally believed to be a solvent-induced entropic
effect $^{[15]}$.  Hydrophobic forces are generated
by the spatial variation of the solvent free energy and not the
enthalpy alone (as with, say, van der Waals forces). These hydrophobic
forces are here considered ``effective'' in the sense that
explicit solvent molecules are not included in their
calculation.

\vspace{5mm}
It is widely assumed that the ``hydrophobic effect'' occurs
because nonpolar solutes do not
participate significantly in hydrogen bonding.  The
presence of these nonpolar solutes in a polar liquid such
as water thus disturbs the local hydrogen bond structure
and leads to an attractive effective interaction between the nonpolar
portions of the solute molecules.  The hydrophobic effect is
usually taken to be proportional to the solvent-accessible
surface area of the nonpolar molecules.

\vspace{5mm}
Such a concept is difficult to employ directly within a numerical
simulation, for these calculations rely upon knowing the
interactions between point particles, rather
than between surface areas.  Since the presence of carbon atoms
is generally the defining feature of nonpolar solutes, it is
therefore sensible to define the carbon atoms in hydrophobic
residues as the centers of the hydrophobic interaction. The
consistency of this approach can be tested simply by
plotting the transfer free energy of hydrophobic amino
acids between polar and nonpolar solvents {\em as a function
of the number of side-chain carbon atoms}.  If the hydrophobic
effect is essentially produced by carbon atoms, and if, in
addition, all carbon atoms in hydrophobic residues contribute
roughly equally, a linear plot will result.  This result
is also implied if the carbon atoms are taken as hard spheres
of excluded volume which interact with water independently of one another.

\vspace{5mm}
A canonical example is shown in Fig. 1.  In his classic analysis
of the hydrophobic effect, Chothia $^{[16]}$ plotted transfer data
of Nozaki and Tanford $^{[17]}$ versus the accessible surface area.
The data could be described by two lines, for nonpolar and somewhat
polar residues, respectively.  By contrast, a plot of the same data versus
the number of side-chain carbon atoms suggests
a fit to a {\em single} line.

\vspace{5mm}
Other hydrophobicity data $^{[18-21]}$ were also analysed in this
fashion.  It was found that the data were reasonably described
by a linear fit to {\em either} the accessible area
{\em or} the number of side-chain carbon atoms.  For long
alkane chains, the two descriptions are essentially
identical.  The slope of the linear fit of free energy
versus number of side-chain carbons, $\Delta$V, ranges from
300 cal/mol$\cdot$carbon to 700 cal/mol$\cdot$carbon,
depending upon the specific points used and the estimated
errors.  In this work, the midpoint value

\begin{equation}
\Delta V = 500 \,\,\, cal/mol\cdot carbon
\end{equation}

\noindent
is used.

\vspace{5mm}
Experimental work on the hydrophobic force suggests that it
decays exponentially with distance.  For example, in their
ground-breaking study Israelachvili and Pashley $^{[22]}$
reported an exponential force with a decay length of 10 {\AA}.
(This value for the decay length was subsequently employed
in the folding study of myoglobin by Saito et.al. described
above $^{[5]}$.)
It is important to note, however, that this experiment
never directly observed an attractive force.  Instead, a
theoretical result was subtracted from the observed
repulsion to estimate the attractive hydrophobic effect.
Later experimental work $^{[23-24]}$ reported decay lengths
of 20 to 160 {\AA}.  A critical discussion of this work
was given by Podgornik and Parsegian $^{[25]}$.

\vspace{5mm}
It seems unlikely that an interaction whose decay length
is of the order of 100 {\AA} could be relevant for
folding myoglobin, which is essentially an oblate
spheroid, 44 x 44 x 25 {\AA} $^{[26]}$.  A sensible
question to ask is what the appropriate decay length should
be for the hydrophobic interaction which folds myoglobin.
Presumably, the hydrophobic effect derives from the local
disruption of a partially-ordered water structure.
An estimate of the decay length should then be possible from the
properties of liquid water alone $^{[27-31]}$.  The average
distance between molecules in ice is 2.8 {\AA}, which
also approximates the location of the first peak in the radial
distribution function g(r) for liquid water.  Estimates
$^{[32]}$ of the thickness of the hydration shell around
typical proteins also yield 2.8 {\AA}.  Moreover,
solvent-accessible area of nonpolar solutes is typically determined
$^{[21]}$ by using a sphere of diameter 2.8 {\AA}. It is, therefore,
reasonable to take
the value $\lambda$ = 2.8 {\AA} assumed here for the
decay length of the hydrophobic force.

\vspace{5mm}
For the reasons discussed above, carbon atoms in
hydrophobic residues are taken as the centers of the
hydrophobic interaction.  Two carbon atoms in
separate hydrophobic residues will then contribute an
amount

\begin{equation}
V_H(r) = V_H(0) exp(-r/\lambda)
\end{equation}

\noindent
to the free energy.  Here, $r$ represents the distance
between the carbon atoms.  It remains to specify the
interaction strength $V_H(0)$.  Recall from Eq. (1)
that the typical free energy of transfer of a
hydrophobic carbon from a nonpolar solvent to a
polar solvent was estimated as $\Delta$V = 500 cal/mol.
This may also be taken as the amount by which the free
energy decreases as a carbon atom approaches a hydrophobic
center.

\vspace{5mm}
The parameter $V_H(0)$ can then be fixed as follows:
The Lennard-Jones interaction between two atoms
can be specified in terms of its
value $E_m$ at its minimum $r_m$

\begin{equation}
V_{LJ} (r) = E_m [-(\frac{r_m}{r})^{12} + 2(\frac{r_m}
{r})^{6}]
\end{equation}

\noindent
The parameterization of Levitt $^{[33]}$ yields the values
$E_m = - 0.19$$\;\;$kcal/mol
at $r_m = 3.53$ {\AA} between
aliphatic carbon atoms.  The minimum of the potential
$V_{LJ}(r) + V_H(r)$ should be less than $E_m$ by an
amount $\Delta$$V$ [see Fig. 2a].  This requirement determines
$V_H(0)= - 1.65$ kcal/mol, which completes the specification
of the hydrophobic force.

\vspace{5mm}
The model proposed in Eq. (2) and the physical arguments leading
to the choice of parameters used can also be justified independently
by making a comparison with the microscopic theory proposed
by Pratt and Chandler $^{[34]}$.  The theory of Pratt and
Chandler involves microscopically well-defined approximations
based upon a hard-sphere treatment of fluids, and is therefore
philosophically consonant with the present model (see the discussion
of Fig. 1 above).  Moreover, it agrees remarkably well quantitatively
with subsequent computer simulations using realistic models
of water (see, e.g., $^{[35-37]}$ and references cited therein).

\vspace{5mm}
The model of Pratt and Chandler for a hard-sphere radius of
3.5 {\AA} (which is the minimum of the Lennard-Jones
potential given above) at 25$^o$C is compared with the present model
in Fig. 2b. The two potentials clearly agree well in terms
of magnitude and range.  One significant difference does,
however, exist.
Superimposed upon the anticipated exponential decay
are a series of potential energy
barriers which oscillate with a period of roughly 3 {\AA},
essentially equal to the hard-sphere diameter of water.
These barriers arise from the existence of water
molecules between the separated hydrophobic pair.  As
the hydrophobic molecules approach one another, the
intervening hydration structures will be altered, and
the water molecules between the hydrophobic
molecules will be forced away.  The effective
potential which determines the interaction must, therefore, change
as the hydrophobic molecules approach one another,
and the relevant energy barrier will disappear.
Dynamical effects of this nature
presumably occur on a picosecond time scale, while
for the present study
(which uses a {\em step size} larger than ten picoseconds)
only processes on the level of nanoseconds to microseconds
are relevant.
The effective
hydrophobic potential model Eq. (2)
is intended to represent a long-time average of the entropic solvent
forces active during folding, and so the short-time
energy barriers are omitted.
This is probably about
as good as one can do at present without
further assumptions or
some remnant of dynamic water molecules.
(This matter is discussed further in the
concluding section).

\vspace{5mm}
It is well to note one final point.  The model of Pratt
and Chandler is valid in the limit of a dilute solute.
Moreover, the present approach treats each hydrophobic
atom as acting independently of other hydrophobic
atoms.  This would not be the case if significant collective
hydrophobic effects were involved.  However, the transfer
free energy of pure hydrocarbon chains is linearly proportional
to the number of carbon atoms in the chain $^{[21]}$,
and the free energy of transfer is generally considered
to be linearly proportional to solvent-accesible surface area,
implying that collective
effects (which would produce nonlinear behavior) are
insignificant.

\vspace{5mm}
b.  Definition of the potential model.

\vspace{5mm}
The protein studied here is sperm-whale myoglobin (1MBN), with
coordinates taken from the Brookhaven Data Bank $^{[38]}$. The
protein is separated into the traditional 8 helices, with
residue assignments as follows:  $A$(3-18), $B$(20-35), $C$(36-42),
$D$(51-57), $E$(58-77), $F$(86-94), $G$(100-118), and $H$(124-149).
Each helix is viewed as a fundamental unit.  All atoms
in each helix are moved as a unit, and only forces between
atoms in different helices are considered.  A ``spring''
connects the final alpha carbon of each helix with the initial
alpha carbon of the next helix.  The spring potential is
taken to be

\begin{eqnarray}
V_{spring} &=& f(r-r_0) \,\,\,\, kcal/mol \nonumber\\
f(x) &=& 5 x ^2 \,\,\,\, x > 0 \nonumber\\
&=& 1 x ^6 \,\,\,\, x < 0
\end{eqnarray}

\noindent
where $r_0$ is set to $3.80 + 1.35 N_{res}$ angstroms, with
$N_{res}$ the number of intervening amino acid residues.
As can be seen in Fig. 3, this
potential allows about an angstrom of free motion about
its minimum, consistent with the notion that the spring
represents a segment of random coil configuration.

\vspace{5mm}
The computation time required to determine forces
between $N$ particles is roughly proportional to $N^2$.
It is therefore important
to eliminate interactions whose contribution
to the free energy is small.  In the parameterization
of Levitt $^{[33]}$ used above, the interactions between
carbon atoms provide the largest contributions by far.
Accordingly, only interactions between helical carbon atoms are
included, with the parameters above used throughout.
Lennard-Jones forces between all helical carbon atoms are included.
Hydrophobic forces [from Eq. (2)] between helical carbon
atoms in the residues, TRP, PHE, TYR, ILE, LEU, VAL,
and MET are also included.  The heme group is
considered as part of the $F$ helix, and all of its
carbon atoms are taken as hydrophobic.

\vspace{5mm}
Some electrostatic effects are included implicitly by
considering the hydrophobic effect to be operative
only between the residues in the above set.  (This
is equivalent to assuming that, for large-scale folding,
the dominant contribution of electrostatic effects is
to reduce or eliminate hydrophobicity).  The
contribution of main-chain carbon atoms to the folding
pattern was also found to be negligible.

\vspace{5mm}
c.  Solving the model:  Brownian dynamics and the
Langevin equation.

\vspace{5mm}
Numerical simulations based upon the Brownian dynamics
of a Langevin equation are indicated when motions on
widely separated timescales are involved $^{[39,40]}$.
A protein in a water solvent is a prime example of this
phenomenon.  Large numbers of rapidly moving solvent
molecules are present and slowly cause the protein to
fold.  Present computational means are insufficient for
the inclusion of explicit solvent molecules, which are
inherently of no direct interest.  It is, therefore,
appropriate to approximate water solvent effects by the
Langevin formalism in conjunction with the hydrophobic
forces defined above.

\vspace{5mm}
The Langevin formalism is also appropriate for another
reason.  A system which evolves by Brownian dynamics
will, in general, achieve thermal equilibrium $^{[41,42]}$.
In thermal equilibrium, the most probable state
corresponds to the global minimum of the free energy,
which here is the native state of the protein.  The
model protein will, therefore, always reach its native
state.  Langevin dynamics thus possesses a distinct
advantage over methods like gradient minimization
or energy-conserving molecular dynamics, neither of
which is guaranteed to find the native state.
Moreover, the physical derivation of the formalism
suggests that it will reach the native state in a
fashion similar to the way a real protein folds
{\em in vitro}.

\vspace{5mm}
In the Langevin approach, solvent effects are
mimicked by the inclusion of random thermal forces
and viscous friction $^{[43]}$.  For a single particle,
the Langevin equation is

\begin{equation}
m \frac{d {\bf v}}{dt} = {\bf F} - \alpha {\bf v} +
{\bf F}^{\prime}(t) \: ,
\end{equation}

\noindent
where $m$ and {\bf v} are the particle mass and velocity,
{\bf F} is the external force as derived from a potential, and
{\bf F}$^{\prime}$ is the fluctuating thermal force.  For a spherical
particle of radius {\em a} in a liquid of viscosity $\eta$,
Stokes' law yields $\alpha$ = 6$\pi$$\eta$${\em a}$.
The random force {\bf F}$^{\prime}$ is specified by its ensemble
averages:

\begin{eqnarray}
<{\bf F}^{\prime}(t)> &=& 0 \nonumber\\
<F_a^{\prime}(t_1)F_b^{\prime}(t_2)> &=& 2\:\:\delta_{ab}\:\:kT\:\:
\alpha\:\:\delta(t_1-t_2)
\end{eqnarray}

\noindent
where T is the temperature.

\vspace{5mm}
Here the Langevin approach is employed to generate
representative trajectories for the eight alpha helices
of myoglobin.  Each helix is treated as a rigid body, with
three translational and three rotational degrees of freedom.
Great simplification of the Langevin equations results if
atoms (except hydrogen, which is ignored) are considered
to be identical spheres, with equal average mass and
viscosity.  (This approximation should only affect the
dynamics slightly and has no effect on the location of
the potential minimum).  Then the Langevin equation
for the center-of-mass coordinate of each helix is
formally identical to Eq. (5), with a suitable
redefinition of variables.

\vspace{5mm}
The rotational kinematics of the helices is also then
straightforward.  Each helix possesses an angular
momentum {\bf L} which evolves in time according to

\begin{equation}
\frac{d {\bf L}}{dt} =\mbox{\boldmath$\tau$} - \frac{\alpha}{m}
\,\,\,{\bf L} + \mbox{\boldmath$\tau$}^{\prime}
\end{equation}

\noindent
in the space-fixed system.  Here, \mbox{\boldmath$\tau$} is the
torque on the helix arising from the potential-derived
force, and \mbox{\boldmath$\tau$}$^{\prime}$ is the torque from random
thermal (solvent) motion.  The ensemble averages
corresponding to Eq. (6) are easily seen to be

\begin{eqnarray}
< \mbox{\boldmath$\tau$}^{\prime}>  &=& 0  \nonumber\\
<\tau^{\prime}_{a}(t_1) \tau^{\prime}_{b}(t_2)> &=& 2
kT\;\;\;\frac{\alpha}{m}\;\;\; I_{ab}\;\;\; \delta(t_1 - t_2)
\end{eqnarray}

\noindent
where $I_{ab}$ is the inertia tensor for particles
of equal mass $m$.

\vspace{5mm}
The translational degrees of freedom are easily
integrated using a velocity Verlet algorithm $^{[39]}$,
but the rotational degrees of freedom require special
techniques based upon quaternionic methods $^{[44,45]}$.
The solvent-induced fluctuating thermal forces are
calculated at each time step by a Gaussian random
number generator with appropriate variance.

\vspace{5mm}
The usual checks of energy conservation (in the absence of
thermal forces and friction) and equipartition were made.  The time
step size was continuously varied according to the
convergence of the algorithm, with a minimum step size
of about ten picoseconds.  The longest time scale which
could be considered is
of the order of
hundreds of microseconds.  Although this is much
larger than the 130 picosecond time scales considered in
classic molecular dynamics simulations $^{[6]}$, it is still
insufficient to probe true protein folding scales, which
are in the domain of seconds to minutes.  An
additional artifice was therefore employed.  The temperature of
the system was initially set to several thousand degrees,
and it was gradually cooled to 25$^o$C.  Energy barrier
penetration thus occurred at a much faster rate than
at normal temperatures, and the folded state was
reached fairly quickly.  This heating/cooling scheme
resembles  Monte Carlo
``simulated annealing'' methods in common
use, but differs in that a realistic continuous folding
trajectory is produced here.  This trajectory might
therefore be similar to the true refolding pathway of the protein.

\vspace{5mm}
\noindent
3.  Results:  A time series of folding myoglobin.

\vspace{5mm}
Figure 4a shows a plot of the kinetic and potential
energy of the model as
it evolves through a typical folding sequence.  Two lines
are plotted, one depicting the sum of the hydrophobic and
Lennard-Jones potential terms, and one which includes
in addition the helix kinetic energy.
Five points are singled out, beginning
with the initial configuration ${\em A}$ and continuing to
the final point ${\em E}$.  The protein configurations
corresponding to these points are shown in Figs. 5a
through 5e, respectively. Figure 4b shows the corresponding
rms deviation (defined below), and the radius of gyration
(using equal masses for all helical atoms).  Both are
plotted as a function of time.

\vspace{5mm}
In this sequence, the initial unfolded state undergoes
a rapid collapse to a folded state near the native
conformation.  Initially, the temperature was set to
2000$^o$C; the system was cooled to 25$^o$C over the
course of 300 nanoseconds.  The viscosity parameter
was adjusted to ensure optimum convergence, resulting
in a thermal relaxation time of about 100 picoseconds.
This run required about 10 hours on a Silicon Graphics
R4000 Iris Indigo.
At the end of the run, the system was reheated and
recooled six times (after the fashion of
simulated annealing).  The protocol was as before,
with the state of lowest energy saved at the end
of each run.  The protein settled
to what appears to be the unique ground state, with the
combined potential from Lennard-Jones and hydrophobic effects
equal to -1000 kcal/mol.  This state is drawn in
Fig. 6a.  For comparison, the native protein is shown
in Fig. 6b.  The rms difference between the two
(taken over all helical nonhydrogen atoms) is 5.95 {\AA}.
The specific definition $^{[46,47]}$ of rms difference is that used
by Cohen, Richmond, and Richards $^{[4]}$ in their classic
work discussed above:

\begin{equation}
\overline{(\Delta r^2)}^{\frac{1}{2}} = \surd{\frac{3}{2}}
[{\frac{2}{(n-1) (n-2)}} \sum^n_{i=1} \sum^{i-1}_{j=1}
(d^{\exp}_{ij}-d^{obs}_{ij})^2]^{\frac{1}{2}}
\end{equation}
wherein consideration of helix pairing
sites leads to a set of 20 possible structures for myoglobin.
Of this set, the structure which most resembles native myoglobin
has an rms deviation of 3.62 {\AA} from the native structure.
It must be pointed out, however, that this structure was
selected from the final set of 20  by visual inspection, rather than
by their helix pairing criteria.  The worst rms deviation
of their 20 proposed structures was 7.6 {\AA}.

\vspace{5mm}
An explicit comparison of the model and native structures is shown in
Fig. 7.  Distances between alpha carbons were determined
for all helical residues in the model and native
conformations.  Contoured areas indicate distances of
less than 13 {\AA}.  The hydrophobic helix-helix
intersection points (BE, BG, AH, FH, and GH) used in
[4] to derive the structure of myoglobin are explicitly
indicated.  Except for the relatively unimportant FH
contact, agreement is fairly good.  (It should be
pointed out that the CD region is fairly nonhelical,
and, as expected, produces the main difference between
the model and native structures.)

\vspace{5mm}
In order to search for other states of
lower energy, five hundred other runs with very
different starting configurations were performed.
No states of lower energy than that in Fig. 6a were found,
and the minimum energy configuration appears to be unique.
Typically, there
was an initial rapid inertial collapse, followed by a subsequent
(slow) stochastic refinement of structure.  This behavior has also
been observed in lattice models of protein folding
$^{[48]}$.  Often, during the initial collapse phase, a
helix becomes trapped between two other helices,
resulting in a collapsed state of higher potential energy
than the minimum.  This defect generally disappears during
subsequent annealing runs, and the system then relaxes
to the true minimum.
Interestingly enough, the inertial collapse often
leads to a structure which has almost
the correct topology of the native state.
During the collapse, the protein dynamically tunnels through
energy barriers, effectively
making ``choices'' about which folding pattern to
adopt. Some of these choices are quite poor, and lead
to helices being trapped by other helices.
One example of this behavior is shown in  Fig. 5e,
where the F helix is caught behind the H helix, and
so encounters an energy barrier on the way to the
folded state.  As can be seen from Fig. 4a, this
state is higher in energy than the minimum state by roughly 200
kcal/mol.

\vspace{5mm}
The  {\em in vitro} folding of myoglobin is not yet completely
understood, so a detailed comparison of the folding pathway
shown here with experiment is impossible.  However, some results
have been reported.  When the heme group is removed from
native myoglobin to produce apomyoglobin, the native state is
destabilized, and a stable intermediate appears in low
pH-induced folding $^{[12]}$.  The detailed three-dimensional
structure of apoMb is not known, but it has been proposed
$^{[13,49-51]}$ that the A, G, and H helices fold to a structure
of moderate stability which constitutes an intermediate state
of the pathway of apoMb folding.  Further work on apoMb $^{[52]}$
finds that stabilized secondary structure appears in the A, G,
and H helices as a compact folding intermediate in less than 5 ms.
By contrast, refolding experiments on intact native myoglobin
$^{[8-11]}$ find no evidence for a stable folding intermediate.
It is suggestive to compare the time series
displayed in Figs. 5 (note especially Fig. 5d), where the
AGH triplet reaches the topology of the native state considerably
in advance of the B-E subdomain.
Nevertheless, the work reported in this paper concerns only intact
native myoglobin, so such a direct comparison with apoMb, though
encouraging, is tenuous.

 \vspace{5mm}
It has also been found $^{[51]}$ that the H helix of intact
myoglobin spontaneously populates helical conformations
irrespective of the state of folding of the remainder of the
polypeptide chain.  These authors suggest that this helix
can be considered an autonomous folding unit.  Additionally,
a G helix peptide segment, though unfolded in aqueous solution $^{[51]}$,
forms an ordered helix in TFE (2,2,2-trifluroethanol) $^{[53,54]}$.
It was proposed that in this case TFE might model the effects
of stabilizing tertiary contacts.

 \vspace{5mm}
Taken as a whole, these data are consonant with the philosophy
that myoglobin folds by the coalescence of nascent metastable
substructures $^{[7]}$ (e.g., ``molten globules'' $^{[55,56]}$).
These subunits do not possess a fixed spatial conformation,
but do on average exhibit a high degree of secondary structure.
As folding progresses, the substructures are stabilized by
tertiary interactions and presumably become the familiar
alpha helices.  Most likely, the effects of this partition
of configuration space manifest themselves early in the
folding process.  The experimental results reviewed above
suggest that this reduction of the relevant degrees of
freedom in the system is completed
in time scales of less than a few milliseconds.
It may thus be a reasonable approximation to model the
nascent subunits by native alpha helices.

 \vspace{5mm}
One important test of this philosophy can be performed
in the present model.  If each helix in the model forms
a realistic approximation to a nascent subunit, then fluctuations
expected to occur during the folding process should not
cause a significant change in the final folded structure.
An additional check of the model was therefore made.  In
a real protein, the amino acid side-chains are in
constant motion.  Presumably, they attain their
native configurations only ${\em after}$ the
initial rapid collapse.  It follows that an initial state
with a different configuration of rotamers should
fold to nearly the same state as obtained with
native rotamer helices.  This expectation was tested
by replacing each amino acid side-chain by the most
probable rotamer state and repeating the calculation.
The resulting minimum is displayed in Fig. 8.  As can
be seen, only slight differences result.  This state
has Lennard-Jones plus hydrophobic potential energy equal
to -933 kcal/mol, and an rms difference of 6.28 {\AA}
from the native conformation, using the same definition $^{[46,47,4]}$
as before.  The ``rotamer'' configuration has an
rms difference of 5.56 {\AA} from the model configuration
Fig. 6a.

\vspace{5mm}
The model can be tested further by applying it to the
protein leghemoglobin.  Leghemoglobin is a protein of
153 residues, of which only 10 are common to the 153
residues of sperm-whale
myoglobin.  The biological origin of leghemoglobin,
a single-chain plant protein which binds oxygen in legume root nodules,
is also very different from that of sperm-whale
myoglobin.  Nevertheless, the native structure of
these two proteins is noticeably similar.
Application of the folding model outlined
above to leghemoglobin accordingly provides an unusually stringent test
of the above model, for it is probably the most dissimilar
protein one can find which possesses a closely related folding
pattern.

\vspace{5mm}
The leghemoglobin coordinates $^{[57]}$
used for this test (2LH7) were obtained
from the Brookhaven Data Bank.
The protein was separated into the traditional 8 helices, with
residue assignments as follows:  $A$(4-19), $B$(21-36), $C$(37-43),
$D$(52-58), $E$(59-78), $F$(87-98), $G$(104-120), and $H$(127-153).
Other details of the potential model were as before.  The
unique minimum energy configuration is displayed in Fig. 9a,
and the native state in Fig. 9b.  The minimum energy
configuration had an energy of -982 kcal/mol, and an
rms deviation of 5.14 {\AA} from the native structure.
The EFGH region of the minimum energy configuration
is quite reasonable, and the topology of the globin fold is manifest.
In a fashion similar to
the case of sperm-whale myoglobin, the partly helical CD
region is somewhat poorly represented. As discussed further
in the next section,
the model also overestimates the attractive interaction
between the heme group and the pair of PHE residues 29 and 30
in the center of the B helix.
This causes the B helix to be pulled
in closely to the heme group and, in turn, forces an additional
misalignment of the A helix.  Overall, however, the
helices do fold to roughly the proper locations,
and so the desired ``sketchbook'' accuracy is obtained.

\vspace{5mm}
\noindent
4.  Conclusions.

\vspace{5mm}
Protein folding presents a challenge of monumental proportions.
Despite the large amount of attention this problem has
received, the physical principles which determine global
folding patterns are largely unknown.  Although sophisticated
pattern recognition techniques may be useful for an initial
survey of this unmapped territory, true understanding must
be based upon the development and testing of physical models
of large-scale folding.

\vspace{5mm}
A small step in this direction was taken here.  The fundamental
philosophy was that globin folding proceeds
by the coalescence of nascent substructures, approximated
here as native alpha helices.  It was additionally assumed
that solvent effects determine the large-scale
folding of these proteins.  Accordingly, a quantitative
model of hydrophobic forces was developed
from physical considerations of transfer
free-energy data and shown to agree quantitatively with microscopic
results from hard-sphere models.
The model was employed
here in conjunction with explicit (Lennard-Jones) van der
Waals forces.  Some electrostatic effects were included
implicitly as well.
Simulation of folding was performed
by Langevin techniques crafted to reproduce closely the
kinetic effects of solvent molecules, while maintaining
close contact with classical molecular dynamics.

\vspace{5mm}
The model was applied to two proteins of radically
different amino-acid sequence and biological origin:
sperm-whale myoglobin and leghemoglobin.
In each case, a unique minimum was found and was shown
to correspond well with the topology of the native structure of
the protein.
The model is very economical, and thus allows an extensive
search of configuration space to be performed.
In order to increase the speed of computation,
all carbon atoms were treated
equally.  This is somewhat unrealistic, for
the hydrophobic interaction produced by a carbon
atom in an aromatic ring must be different from
that produced by a carbon
atom in an alkane chain.  Thus, the hydrophobic effective
potential of PHE is overestimated, while that of
LEU and ILE is underestimated.  Nevertheless, the gross
stuctures obtained do, in fact, have the proper topology--the
helices appear in their proper relative
locations. This success implies that many of the
physical principles relevant for large-scale folding
have been accurately captured in the model.

\vspace{5mm}
Once a structure with ``sketchbook'' accuracy is produced,
it is of course possible to include more precise interactions
for a further refinement of the prediction.
This should
be computationally feasible when folding is nearly
complete and the gross structure of the protein has been obtained.
The present model, however, does allow
the large-scale folding process to be observed directly,
which should be useful for further design of physical
models and algorithmic developement.
In particular,
the model exhibited a number of interesting phenomena supportive
of the concept $^{[58]}$ that protein folding is ${\em not}$
simply a uniform collapse, simultaneous on all
length scales.  This may be heartening news
for the protein folding problem in general, for it
suggests that not all interactions are relevant at
all times during the folding process.  It may therefore
be appropriate to address protein folding by developing
a hierarchy of ``effective'' interactions between
subunits of various sizes.  It is, in a sense,
equivalent to determining the correct ``language''
for the description of protein folding (one does not generally
use subatomic physics to describe chemistry).
Significant simplification could then well occur.

\vspace{5mm}
The physical principles which underly this point
are quite clear, and can be illustrated by an elementary paradigm.
Consider an initial configuration of the model where
the helices are all aligned in a
straight line.
All eight helices will experience torques which cause them
to rotate toward the configuration of least energy.
The A and H helices are initially
the least constrained, for each of the remaining helices
is connected at both ends to other helices
by a ``string'' of random coil.
It is energetically
most favorable either to move only these two helices,
or else to move ${\em groups}$ of helices which include either the
A or H helix.  (This argument also suggests that refolding
for an alpha-helix protein like myoglobin should begin
at the ends of the molecule, which might partly explain
the early importance of the A, G, and H helices in refolding
processes $^{[49-54]}$).
As a consequence, typically groups of several helices must move
together as subunits.
Collective coordinates even larger than alpha helices therefore
are important role at certain stages of the
folding process, and their appearance
offers additional validation of the idea that the
identification of relevant large-scale degrees of freedom
(such as the alpha helices taken here as a basis set)
is important in protein folding.
If this spontaneous reduction of the degreees of freedom
in the system is a general physical principle,
it could provide one possible resolution to
the Levinthal paradox $^{[2]}$ mentioned in the Introduction.
Thus, if the number of physically relevant degrees of freedom in the system
is far fewer than those present in the entire configuration space, a
considerably faster search for the folded state is implied.
This simplification is also useful computationally, for
it suggests that
the speed of folding algorithms may generally be improved
by the inclusion of moves of collective coordinates,
of which nascent alpha helices are only one example.

\vspace{5mm}
Another useful point
is the clear existence $^{[48]}$ of two stages in
folding--an inertial collapse phase followed by
stochastic refinement of structure.  This separation
suggests that substantial inertial effects should be
included in a computationally effective protein
folding algorithm.  Indeed, when the present Langevin
methodology (which includes dynamical inertial effects)
was compared with standard Monte
Carlo simulated annealing, it was found to be
quite superior in performance.  This difference
in performance persisted even when the step size
for individual Monte Carlo moves was allowed to vary dynamically
so as to ensure an acceptance rate of about one half,
in accordance with standard techniques of simulated annealing.
The need for alternatives to pure canonical ensemble Monte Carlo methods
in molecular simulations $^{[59,60,35]}$ and when
short-range forces lead to large-scale collective
effects $^{[61]}$ is well-known.

\vspace{5mm}
Since the present algorithm is based upon classical
molecular dynamics, it uses a real physical time coordinate.
Thus the time taken for various subprocesses can be
estimated from observation of the folding process.
Observation times are computationally limited at present to microsecond
time scales, which requires unphysically high temperatures
to be used in the slow ``stochastic'' second phase of
folding.  However, the initial rapid inertial collapse
was generally visible, even at room temperature.
This suggests that the initial collapse phase typically occurs
in a microsecond time scale, which is also the time scale
at which alpha helices form and disappear $^{[62,63]}$.
Therefore the inertial collapse occurs at the same
time that the alpha helices themselves are becoming stable.
One implication of this result is that alpha helices
become a significant part of the protein folding pathway
before they are actually stable entities.  As has been
pointed out previously $^{[63]}$, there is no meaningful
threshold of stability for alpha helices
to become relevant in folding.

\vspace{5mm}
One of the virtues of the present model is its economy,
which permits an extensive search of configuration space
to be made with present
computational architecture.
Although more detailed models of the effective
hydrophobic interaction can be constructed (see, e.g., $^{[64]}$),
their complexity is such that the accessible time
scales are typically of the order of hundreds of picoseconds.
Explicit inclusion of water solvent molecules reduces
this accessible timeframe further, by another factor
of twenty $^{[64]}$.
The philosophy presented here allows effects
whose natural timescale is of the order of tens to hundreds
of microseconds to be considered, albeit with generally less
precision.  The accessible timescale is, however,
increased by a factor larger than a million.

\vspace{5mm}
A natural question to ask, therefore, is
whether the model can be simplified further.  One
attempt in this direction was tried.  The attractive portion
of the Lennard-Jones potential was removed, and the coefficient
of the hard-core repulsion was varied to give different
values of the minimum hard-sphere distance of the potential (which is
then the only parameter remaining, aside from the size of water
used as the decay length).
The resulting model is then probably as simple
as can be conceived, for a repulsive part of the
potential is required to prevent the helices from
collapsing into one another, while an attractive
force is needed to drive the folding process.
Reasonable folded
structures could be obtained when this minimum value
was taken to be 4.0 {\AA},
which occurred when the repulsive part of the
potential was the same as employed in the full model.
The rms value so obtained was
6.07 {\AA}, to be compared with 6.59 {\AA} and 6.61 {\AA}
reached for potential minima of 3.5 {\AA} and 4.5 {\AA}
respectively.  Values of the minimum significantly less
than 4.0 {\AA} led to overly compact structures, while
larger values produced structures which were too extended.

\vspace{5mm}
The fact that somewhat plausible structures can be obtained
when the only force driving folding is the hydrophobic
effect provides {\em quantitative} confirmation of the idea
that it is this effect which determines large-scale globin folding
patterns.  There is a philosophical inconsistency in removing the
attractive part of the Lennard-Jones potential, however, for
the hard-sphere radius of a molecule is typically taken
as the location of the minimum of this potential.
When the attractive
part of the Lennard-Jones force is eliminated,
the Lennard-Jones potential then has no minimum,
and the hard-core radius is undefined.
To prevent collapse, the attractive hydrophobic force must be
balanced by an arbitrarily chosen hard-core repulsion.
The hard-core repulsion then must be parameterized and
determined by observing the folding of myoglobin or
some similar process.
By contrast, the hydrophobic
potential model applied in this work was derived
from physical arguments,
independently of knowledge of the structure of myoglobin.
The recognizable structure so obtained accordingly
suggests that these arguments have a fair degree of validity.

\vspace{5mm}
In summary, then, the purpose of this work is not to
attain the most {\em precise} depiction of folding
possible, but rather to attempt to learn the most
{\em useful} language in which to describe the process.
With few exceptions, all useful descriptions of complex
systems involve judicious approximations.  So it is here.
A careful attempt was made to develop an accurate
physical model of the most important effects which
determine folding patterns.  Rather than taking an
intricate potential function and determining its
parameters empirically (ultimately equivalent to a form of
pattern recognition), the goal of logical simplicity
was sought throughout.  The result is a tractable model
of protein folding based upon sound physical principles
which, nevertheless, captures many of the essential
features of more detailed models.  Extension of the
present formalism to other proteins should, therefore,
be of some utility in the systematic extraction of the
physical principles which determine large-scale folding
patterns.

\newpage
\noindent
\underline {Acknowledgements}.

\vspace{7mm}
It is a pleasure to thank S.K. Burley, B. Chait, K. Clark, M.D. Doyle,
N.N. Khuri, B. Knight, J. Kuriyan, J. Lederberg, M. Levitt, P. Model and
D. Zwanziger for useful discussions and technical assistance.  This
manuscript was typeset in LaTeX by Toni Weil.

\newpage
\begin{center}
{\bf References}\\
\end{center}

\begin{enumerate}

\item  Anfinsen, C.B.:  Principles that govern the folding
of protein chains.  Science 181:223-230, 1973.

\item  Levinthal, C.:  Are there pathways for protein
folding?  J. Chim. Phys. 65:44-45, 1968.

\item  Ptitsyn, O.B., Rashin, A.A.:  A model of myoglobin
self-organization.  Biophys. Chem. 3:1-20, 1975.

\item  Cohen, F.E., Richmond, T.J., Richards, F.M.:Protein
folding: Evaluation of some simple rules for the assembly
of helices into tertiary structure with myoglobin as an
example.  J. Mol. Biol. 132:275-288, 1979.

\item  Saito, N., Shigaki, T., Kobayashi, Y., Yamamoto,
M.: Mechanisms of protein folding: I. General considerations
and refolding of myoglobin.  Proteins 3:199-207, 1988.

\item  Kuczera, K., Kuriyan, J., Karplus, M.: Temperature
dependence of the structure and dynamics of myoglobin: A
simulation approach.  J. Mol. Biol. 213:351-373, 1990.

\item  Bashford, D., Cohen, F.E., Karplus, M., Kuntz, I.D.,
Weaver, D.L.: Diffusion-collision model for the folding
kinetics of myoglobin.  Proteins. 8:211-227, 1988.

\item  Acampora, G. and Hermans, J.: Reversible denaturation
of sperm-whale myoglobin.  I.  Dependence on temperature,
pH, and composition.  J. Am. Chem. Soc. 89: 1543-1547, 1967.
II.  Thermodynamic analysis. ibid. 1547-1552.

\item  Shen, L.L., and Hermans, J.: Kinetics of conformation
changes of sperm-whale myoglobin.  I.  Folding and
unfolding of metmyoglobin following pH jump.  Biochemistry
11: 1836-1841, 1972.

\item  Puett, D.: The equilibrium unfolding parameters of horse
and sperm-whale myoglobin.  J. Biol. Chem. 247: 4623-4634, 1973.

\item  Privalov, P.L., Griko, Yu. V., Venyaminov, S. Yu.,
Kutyshenko, V.P.: Cold denaturation of
myoglobin.  J. Mol. Biol. 190:487-498, 1986.

\item  Griko, Yu. V., Privalov, P.L., Venyaminov, S. Yu.,
Kutyshenko, V.P.: Thermodynamic study of the apomyoglobin
structure. J. Mol. Biol. 202:127-138, 1988.

\item  Hughson, F.M., and Baldwin, R.L.: Use of site-directed
mutagenesis to destabilize native apomyoglobin relative to
folding intermediates.  Biochemistry 28:4415-4422, 1989.

\item  Dill, K.A.:  Dominant forces in protein folding.
Biochemistry 29:7133-7155, 1990.

\item  Nicholls, A., Sharp, K.A., Honig, B.: Protein
folding and association: Insights from the interfacial
and thermodynamic properties of hydrocarbons.
Proteins. 11:281-296, 1991.

\item  Chothia, C.H.:  Hydrophobic bonding and accessible
surface area in proteins.  Nature 248:338-339, 1974.

\item  Nozaki, Y. and Tanford, C.: The solubility of amino
acids and two glycine peptides in aqueous ethanol and
dioxane solutions.  Establishment of a hydrophobicity
scale.  J. Biol. Chem. 246:2211-2217, 1971.

\item  Zimmerman, J.M., Eliezer, N., Simha, R.: The
characterization of amino acid sequences in proteins
by statistical methods.  J. Theoret. Biol. 21:170-201,
1968.

\item  Jones, D.D.: Amino acid properties and side-chain
orientation in proteins: A cross correlation approach.
J. Theor. Biol. 50:167-183, 1975.

\item  Wolfenden, R., Andersson, L., Cullis, P.M.,
Southgate, C.C.B.: Affinities of amino acid side chains
for solvent water.  Biochemistry 20:849-855, 1981.

\item  Sharp, K.A., Nicholls, A., Friedman, R., Honig,
B.: Extracting hydrophobic free energies from
experimental data: Relationship to protein folding
and theoretical models.  Biochemistry 30:9686-9697,
1991.

\item  Israelachvili, J. and Pashley, R.: The hydrophobic
interaction is long range, decaying exponentially with
distance.  Nature 300:341-342, 1982.

\item Claesson, P.M., and Christenson, H.K.: Very long
range attractive forces between uncharged hydrocarbon and
fluorocarbon surfaces in water.  J. Phys. Chem. 92:1650-1655,
1988.

\item  Christenson,H.K., and Claesson, P.M.: Cavitation
and the interaction between macroscopic hydrophobic
surfaces.  Science 239:390-392, 1988.

\item  Podgornik, R. and Parsegian, V.A.: An electrostatic-
surface stability interpretation of the ``hydrophobic''
force inferred to occur between mica plates in solutions
of soluble surfactants.  Chem. Phys. 154:477-483, 1991.

\item  Dickerson, R.E., and Geis, I.: ``Hemoglobin: Structure,
Function, Evolution, and Pathology'': Benjamin/Cummings, 1983.

\item  Eisenberg, D., and Kauzmann, D.:  ``The Structure and
Properties of Water''.  Oxford, 1979.

\item  Franks, F. (ed.): ``Water -- A Comprehensive Treatise''.
Plenum Press, New York, 1972.

\item  Stillinger, F.H.: Water revisited.  Science 209:451-457,
1980.

\item  Ben-Naim, A.: ``Hydrophobic Interactions''.  Plenum
Press, New York, 1980.

\item  Tanford, C.  ``The Hydrophobic Effect''.  2nd ed.
Wiley-Interscience, New York, 1980.

\item  Cantor, D.R., Schimmel, P.R.: ``Biophysical Chemistry,
part II.  Techniques for the Study of Biological Structure
and Function''.  W.H. Freeman, San Francisco, 1980. p. 555.

\item  Levitt, M.: Energy refinement of hen egg-white
lysozyme.  J. Mol. Biol. 82:393-420, 1974.

\item  Pratt, L. R., Chandler, D.: Theory of the hydrophobic
effect.  J. Chem. Phys. 67:3683-3704, 1977.

\item  Pangali, C., Rao, M., Berne, B.J.: Monte Carlo simulation
of the hydrophobic interaction. J. Chem. Phys. 71: 2975-2981, 1979.

\item  Jorgensen, W., Buckner, K., Boudon, S., Tirado-Reves, J.:
Efficient computation of absolute free energies of binding by
computer simulations. Application to the methane dimer
in water. J. Chem. Phys. 89:3742-3746, 1988.

\item  Smith, D., Haymet, A.D.J.: Free energy, entropy, and
internal energy of hydrophobic interactions: Computer simulations.
J. Chem. Phys. 98:6445-6454, 1993.

\item  Watson, H.C.:  The stereochemistry of the protein
myoglobin.  Prog. stereochem. 4:299, 1969.

\item  Allen, M.P., Tildesley, D.J.: ``Computer
Simulation of Liquids''.  Clarendon Press, Oxford, 1991.

\item  McCammon, J.A., Harvey, S.C.: ``Dynamics of
Proteins and Nucleic Acids''.  Cambridge, 1988.

\item  Arnold L., Lefever, R.:  ``Stochastic Nonlinear
Systems in Physics, Chemistry, and Biology''.  Springer
Series in Synergetics, v.8.  Springer-Verlag, 1981.

\item  Gardiner, C.W.: ``Handbook of Stochastic Methods for
Physics, Chemistry, and the Natural Sciences''.  Springer
Series in Synergetics, v. 13.  Springer-Verlag, 1983.

\item  Wax, N.:  ``Selected Papers on Noise and Stochastic
Processes''.  Dover, New York, 1954.

\item  Evans, D.J., Murad, S.: Singularity-free algorithm
for molecular-dynamics simulation of rigid polyatomics.
Mol. Phys. 34:327-331, 1977.

\item  Potter, D.: ``Computational Physics''.  Wiley,
New York, 1972.

\item  Nishikawa, A., Ooi, T., Isogai, Y., Saito, N.:
Tertiary structure of proteins. I. Representation and
computation of the conformation.  J. Phys. Soc. Jpn.
32:1331-1337, 1972.

\item  Levitt, M.: A simplified representation of protein
conformations for rapid simulation of protein folding.
J. Mol. Biol. 104:59-107, 1976.

\item  Shaknovich, E., Farztdinov, G., Gutin, A.M.,
Karplus, M.:  Protein folding bottlenecks:  A lattice Monte
Carlo simulation.  Phys. Rev. Lett. 12:1665-1668, 1991.

\item Hughson, F.M., Barrick, D., Baldwin, R.L.:
Probing the stability of a partly folded apomyoglobin
intermediate by site-directed mutagenesis.  Biochemistry
30: 4113-4118,1991.

\item Hughson, F.M., Wright, P.E., Baldwin, R.L.:
Structural characterization of a partly folded apomyoglobin
intermediate.  Science 249: 1544-1548, 1990.

\item Waltho, J.P., Feher, V.A., Merutka, G., Dyson, H.J.,
Wright, P.E.:  Peptide models of protein folding initiation
sites.  1. Secondary structure formation by peptides
corresponding to the G- and H- helices of myoglobin.
Biochemistry 32: 6337-6347, 1993.

\item Jennings, P.A., Wright, P.E.: Formation of a molten globule
intermediate early in the kinetic folding pathway of apomyoglobin.
Science 262: 892-896, 1993.

\item Shin, H.C., Merutka, G., Waltho, J.P., Wright, P.E., Dyson, H.J.:
Peptide models of protein folding initiation sites.
2. The G-H turn region of myoglobin acts as a helix stop signal.
Biochemistry 32: 6348-6355, 1993.

\item Shin, H.C., Merutka, G., Waltho, J.P., Tennant, L.L.,
Dyson, H.J., Wright, P.E.:
Peptide models of protein folding initiation sites.
3. The G-H helical hairpin turn of myoglobin.
Biochemistry 32: 6356-6364, 1993.

\item Ptitsyn, O.B.:  Protein folding: Hypotheses and experiments.
J. Protein Chem. 6: 273-293, 1987.

\item Kuwajima, K.:  The molten globule state as a clue for understanding
the folding and cooperativity of globular-protein structure.
Proteins 6: 87-103, 1989.

\item Arutyunyan, E.G., Kuranova, I.P., Vainshtein, B.K., Steigemann,
W.: Structural investigation of leghemoglobin.  VI. Structure of
acetate-ferrileghemoglobin at a resolution of 2.0 angstroms.
(Russian) Kristallografiya 25: 80, 1980.

\item Jaenicke, R.: Protein folding: Local structures, domains,
subunits, and assemblies.  Biochemistry 30: 3147-3161, 1991.

\item Pangali, C., Rao, M., Berne, B.J.: On a novel Monte Carlo
scheme for simulating water and aqueous solutions. Chem.
Phys. Lett. 55:413-417, 1978.

\item Rao, M., Pangali, C., Berne, B.J.: On the force bias Monte
Carlo simulation of water: Methodology, optimization and comparison
with molecular dynamics.  Mol. Phys. 37:1773-1798, 1979.

\item Callaway, D.J.E., Rahman, A.: Microcanonical ensemble
formulation of lattice gauge theory.  Phys. Rev. Lett. 49: 613-616,
1982.

\item Gruenewald, B., Nicola, C., Lustig, A., Schwarz, G.,
Klump, H.: Kinetics of the helix-coil transition of a polypeptide
with non-ionic side groups derived from ultrasonic relaxation
measurements. Biophys. Chem. 9:137-147, 1979.

\item Creighton, T.E.: Protein Folding.  Biochem. J. 270:1-16, 1990.

\item Stouten, P.F.W., Frommel, C., Nakamura, H., Sander, C.:
An effective solvation term based on atomic occupancies for use
in protein simulations.  Mol. Sim. 10:97-120, 1993.

\item Warshel, A., Levitt, M.: Folding and stability of
helical proteins: Carp myogen.  J. Mol. Biol. 106:421-437.

\item Levitt, M., Warshel, A.: Computer simulation of protein
folding.  Nature 253:694-698.

\end{enumerate}

\newpage
\noindent
\underline{Figure Captions}.

\vspace{.25in}
\noindent
Fig. 1a.

Plot of accessible surface area ({\AA}$^2$) versus
hydrophobicity (kcal/mol) for various amino acid residues.
Lines of slope 26 cal/mol $\cdot$ {\AA}$^2$ and 22
cal/mol$\cdot$ {\AA}$^2$ are also shown.  Adapted from
Chothia [16].

\vspace{.25in}
\noindent
Fig. 1b.

The same data [17] as Fig. 1b plotted versus the
number of carbon atoms in the residue side-chain.  The line
has a slope of 321 cal/mol $\cdot$ carbon.

\vspace{.25in}
\noindent
Fig. 2a.

Solid line displays Lennard-Jones potential
$V_{LJ}(r)$ from Eq. (3).  Dashed line is sum of
$V_{LJ}(r)$ and $V_H(r)$.  Potential minima differ by
500 cal/mol (see text).

\vspace{.25in}
\noindent
Fig. 2b.

Solid line shows the potential of mean force derived
by Pratt and Chandler $^{[34]}$ for a hard-sphere radius of
3.5 {\AA}.  Dashed line shows the hydrophobic effective
potential $V_H(r)$ proposed here.  Energy is plotted in units
of RT at 25$^o$C, distance in {\AA}.

\vspace{.25in}
\noindent
Fig. 3.

Potential function $f(x)$ (kcal/mol) versus $x$
({\AA}) from Eq. 4.

\vspace{.25in}
\noindent
Fig. 4a.

Lower line is a plot of the sum of the hydrophobic and
Lennard-Jones potential terms (kcal/mol) given in the text versus
time (nanoseconds).  Upper
line includes helix kinetic energy.
In addition to the initial state A and the final state E, the
points B, C, and D are labelled.

\vspace{.25in}
\noindent
Fig. 4b.

Plots of rms distance from the native protein (as defined in the
text) and radius of gyration versus time (nanoseconds),
corresponding to Fig. 4a.
The plots are labelled ``rms'' and ``gyr'' respectively.

\newpage
\noindent
Figs. 5a-e.

Myoglobin configurations corresponding to points labelled A
through E in Fig. 4.  Helices are drawn as cylinders along
major axes of inertia tensors.  Heme unit carbon atoms
drawn as spheres, central (larger) atom is iron.  Helix
and atomic radii not to scale.

\vspace{.25in}
\noindent
Fig. 6a.

Unique folded configuration produced by the model, drawn as per
Figs. 5.

\vspace{.25in}
\noindent
Fig. 6b.

Native myoglobin, drawn as per Figs. 5.

\vspace{.25in}
\noindent
Fig. 7.

Plot of distances between alpha carbons of helical
residues.  Axes indicate residue numbers.  Contoured
areas indicate distances less than 13 {\AA}.  Lower
half is native conformation (Fig. 6b), upper half
is model (Fig. 6a).  Hydrophobic contact zones (BG,
etc.) of Ref. [4] labelled as shown.

\vspace{.25in}
\noindent
Fig. 8.

Folded ``rotamer'' state, drawn as per Figs. 5.

\vspace{.25in}
\noindent
Fig. 9a.

Unique folded configuration of leghemoglobin
produced by the model, drawn as per
Figs. 5.

\vspace{.25in}
\noindent
Fig. 9b.

Native leghemoglobin, drawn as per Figs. 5.

\end{document}